\documentclass[aps,pra,showpacs,superscriptaddress,preprint]{revtex4}
\usepackage{amsmath}
\usepackage{graphicx}

\catcode`ð=\active
 \defð{\u{g}}
 \catcode`Ð=\active
 \defÐ{\u{G}}
 \catcode`Ý=\active
\defÝ{\. I}
 \catcode`ö=\active
\defö{\"{o}}
 \catcode`Ö=\active
 \defÖ{\"O}
 \catcode`ü=\active
 \defü{\"{u}}
 \catcode`Ü=\active
 \defÜ{\"{U}}
 \catcode`Þ=\active
\defÞ{\c{S}}
 \catcode`þ=\active
 \defþ{\c{s}}
 \catcode`ý=\active
 \defý{{\i}}
 \catcode`ç=\active
\defç{\d{c}}
 \catcode`Ç=\active
\defÇ{\d{C}}

\begin{document}

\title{Approximate analytical solutions of the generalized Woods-Saxon
potentials including the spin-orbit coupling term and spin symmetry }
\author{Sameer M. Ikhdair}
\email[E-mail: ]{sikhdair@neu.edu.tr}
\affiliation{Department of Physics, Near East University, Nicosia, Cyprus, Turkey}
\author{Ramazan Sever}
\email[E-mail: ]{sever@metu.edu.tr}
\affiliation{Department of Physics, Middle East Technical University, 06800, Ankara,Turkey}
\date{%
\today%
}

\begin{abstract}
We study the approximate analytical solutions of the Dirac equation for the
generalized Woods-Saxon potential with the pseudo-centrifugal term. In the
framework of the spin and pseudospin symmetry concept, the approximately
analytical bound state energy eigenvalues and the corresponding upper- and
lower-spinor components of the two Dirac particles are obtained, in closed
form, by means of the Nikiforov-Uvarov method which is based on solving the
second-order linear differential equation by reducing it to a generalized
equation of hypergeometric type. The special cases $\kappa =\pm 1$ ($l=%
\widetilde{l}=0,$ $s$-wave) and the non-relativistic limit can be reached
easily and directly for the generalized and standard Woods-Saxon potentials.
Also, the non-relativistic results are compared with the other works.

\end{abstract}

\maketitle

\newpage

\section{Introduction}

It is well known that the exact solutions play an important role in quantum
mechanics since they contain all the necessary information regarding the
quantum model under study. The exact solutions of the Schr\"{o}dinger
equation are only possible for a hydrogen atom and for a harmonic oscillator
in three dimensions [1-3]. However, when a particle is in a strong potential
field, the relativistic effect must be considered, which gives the
correction for non-relativistic quantum mechanics [4]. Taking the
relativistic effects into account, a particle in a potential field should be
described with the Klein-Gordon (KG) and Dirac equations. In recent years,
there has been an increased interest in finding exact solutions to Schr\"{o}%
dinger, KG, Dirac and Salpeter equations for various potential schemes
[4--20]. The problems that can be exactly solved for the KG and/or Dirac
equations are seldom except a few examples, such as hydrogen atom and
electrons in a uniform magnetic field. Recently some authors solved such
relativistic equations for some potentials. Ikhdair [6] obtained the
bound-state solution of the $D$-dimensional KG equation for the vector and
scalar general Hulth\'{e}n-type potentials with any arbitrary $l$-state
using the Nikiforov-Uvarov (NU) method. Moreover, E\u{g}rifes and Sever [7]
investigated the bound state solutions of the $1D$ Dirac equation with $%
\mathcal{PT}$-symmetric real and complex forms of generalized Hulth\'{e}n
potential. Yi \textit{et al}. [10] obtained the energy equations in the KG
theory with equally mixed vector and scalar Rosen-Morse-type potentials. We
have solved the spinless $1D$ Salpeter equation analytically for its exact
bound state spectra and wave functions with real and complex forms of the $%
\mathcal{PT}$-symmetric generalized Hulth\'{e}n potential [11]. We obtained
a quasi-exact analytic bound-state solutions within the framework of the
position-dependent effective mass KG equation for scalar and vector Hulth%
\'{e}n potentials in any arbitrary $D$-dimension and with any orbital
quantum number $l$ using the NU method combined with a new approximation
scheme for the centrifugal potential term [20].

The Woods-Saxon (WS) potential and it's various modifications have played an
essential role in microscopic physics in the determination of the energy
level spacing, particle number dependence of energy quantities and universal
properties electron distributions in atoms, nuclei and atomic clusters since
it can be used to describe the interaction of neutron with one heavy-ion
nucleus and also for optical potential model [21,22]. Although the
non-relativistic Schr\"{o}dinger equation with this potential has been
solved for $s$-states [21] and the single-particle motion in atomic nuclei
has been explained quite well, the relativistic effects for a particle under
the action of this potential are more important, especially for a
strong-coupling system. Berkdemir \textit{et al} [22] obtained the
bound-state solution of the $1D$ Schr\"{o}dinger equation for the
generalized WS potential by means of the NU method. We investigated the
bound-state solutions of the $1D$ KG equation with real and complex forms of
the generalized WS potential [12]. Kennedy [23] obtained the scattering and
bound-state solutions of the one-dimensional Dirac equation for the WS
potential. Guo and Sheng [24] solved exactly the $s(\widetilde{s})$-wave
Dirac equation ($l=0,$ i.e., $\kappa =-1$ for spin and $\kappa =1$ for
pseudospin symmetry) for a single particle with spin and pseudospin symmetry
moving in a central WS potential. They obtained the energy spectra of the
bound-states and the corresponding wave functions for the two-component
spinor in terms of the hypergeometric functions. Alhaidari has developed a
new two-component approach in the case of three-dimensional Dirac equation
for the spherically symmetric potential and solved a class of
shape-invariant Morse, Rosen-Morse, Eckart, P\"{o}schl-Teller and Scarf,
potentials and given their rlativistic bound-state spectra and spinor wave
functions [25-27]. Guo \textit{et al} followed Alhaidari's approach and
discussed the Dirac equation with WS and Hulth\'{e}n potentials for
spherical system and given their bound-state spectra and the spinor wave
functions for $s$-states [28,29].

For the more realistic nuclear system where the nucleons are described in
the relativistic mean field with the attractive $S(\overrightarrow{r})$ and
repulsive vector $V(\overrightarrow{r})$ potential, although some numerical
techniques have been developed to Dirac equation, any analytic solution has
not still been obtained for the Dirac-WS problem. In the special cases of
spin symmetry ($\Delta =V-S=A=$cons$\tan $t) and pseudospin symmetry ($%
\Sigma =V+S=A=$cons$\tan $t), Ginocchio \textit{et al} solved triaxial,
axial and spherical harmonic oscillators for the case $\Delta =0$ and
applied it to the study of antinucleons embedded in nuclei [30]. Lisboa
\textit{et al} studied the generalized relativistic harmonic oscillator for
spin $1/2$ particles and obtained the analytic solutions for bound states of
the corresponding Dirac equations by setting $A=0$ [31]. Very recently,
Ikhdair [32] studied the exact solution of the spatially dependent Dirac
equation with the Rosen-Morse potential for arbitrary spin-orbit quantum
number $\kappa .$ Under the conditions of the spin symmetry $S\sim V$ and
pseudospin symmetry $S\sim -V$, the bound state energy eigenvalues and
corresponding upper- and lower-spinor wave functions are investigated in the
framework of the NU method. Furthermore, We have solved the constant mass KG
equation for the Eckart potential [32] and the spatially-dependent mass KG
equation for the Coulomb-like potential [33] and obtained the bound-state
solutions of the energy eigenvalues and wavefunctions.

In this work, we will solve the (3+1) dimensional Dirac equation for a
particle trapped in the spherically symmetric generalized WS potential under
the conditions of exact spin and pseudospin symmetry combined with
approximation for the centrifugal and pseudo-centrifugal terms, and give the
two-component spinor wavefunctions and the energy spectra for any arbitrary
spin-orbit $\kappa $ bound states. The NU method [34] is used in the
calculations. For the $1D$ case, upon changing the values of the potential
parameters from real to pure imaginary, we obtain Hamiltonians that may or
may not be $\mathcal{PT}$-symmetric. In addition, the spinor wavefunctions
and the energy spectra of $s$-wave $(\kappa =\pm 1)$ bound states and the
non-relativistic limit are also discussed for the generalized and for a
special case of the standard WS potential.

The paper is structured as follows: In sect. 2, we outline the NU method and
derive a parametric generalization version. Section 3 is devoted for the
exact analytic bound state energy eigenvalues and two lower- and
upper-spinor components wave functions of the Dirac equation with
generalized WS quantum system obtained by means of the NU method. The spin
symmetry and pseudo-spin symmetry solutions are investigated using the NU
method. In sect. 4, we study the cases $\kappa =\pm 1$ ($l=\widetilde{l}=0,$
$s$-wave) and the non-relativistic limit and compare with other wave
equations and models. Finally, the relevant conclusions are given in sect. 5.

\section{The Nikiforov-Uvarov Method}

The NU method has been used to solve the Schr\"{o}dinger [19], KG [20,33]
and Dirac [28,32] wave equations for central and non-central potentials. Let
us briefly outline the basic concepts of the method [34]. This method was
proposed to solve the second-order linear differential equation of the
hypergeometric-type:
\begin{equation}
\sigma ^{2}(z)u^{\prime \prime }(z)+\sigma (z)\widetilde{\tau }(z)u^{\prime
}(z)+\widetilde{\sigma }(z)u(z)=0,
\end{equation}%
where the prime denotes the differentiation with respect to $z,$ $\sigma (z)$
and $\widetilde{\sigma }(z)$ are analytic polynomials, at most of
second-degree, and $\widetilde{\tau }(s)$ is of a first-degree polynomial$.$
Let us discuss the exact particular solution of Eq. (1) by choosing%
\begin{equation}
u(z)=y_{n}(z)\phi _{n}(z),
\end{equation}%
resulting in a hypergeometric type equation of the form:
\begin{equation}
\sigma (z)y_{n}^{\prime \prime }(z)+\tau (z)y_{n}^{\prime }(z)+\lambda
y_{n}(z)=0.
\end{equation}%
The first part $y_{n}(z)$ is the hypergeometric-type function whose
polynomial solutions are given by the Rodrigues relation%
\begin{equation}
y_{n}(z)=\frac{A_{n}}{\rho (z)}\frac{d^{n}}{dz^{n}}\left[ \sigma ^{n}(z)\rho
(z)\right] .
\end{equation}%
where $A_{n}$ is a normalization factor and $\rho (z)$ is the weight
function satisfying the condition%
\begin{equation}
\left[ \sigma (z)\rho (z)\right] ^{\prime }=\tau (z)\rho (z),
\end{equation}%
with
\begin{equation}
\tau (z)=\widetilde{\tau }(z)+2\pi (z),\tau ^{\prime }(z)<0.
\end{equation}%
Since $\rho (z)>0$ and $\sigma (z)>0,$ the derivative of $\tau (z)$ has to
be negative for bound states [32-34] which is the main essential condition
for any choice of particular solution. The other part of the wave function
is defined as a logarithmic derivative%
\begin{equation}
\frac{\phi ^{\prime }(z)}{\phi (z)}=\frac{\pi (z)}{\sigma (z)},
\end{equation}%
where%
\begin{equation}
\pi (z)=\frac{1}{2}\left[ \sigma ^{\prime }(z)-\widetilde{\tau }(z)\right]
\pm \sqrt{\frac{1}{4}\left[ \sigma ^{\prime }(z)-\widetilde{\tau }(z)\right]
^{2}-\widetilde{\sigma }(z)+k\sigma (z)},
\end{equation}%
with%
\begin{equation}
k=\lambda -\pi ^{\prime }(z).
\end{equation}%
The determination of $k$ is the key point in the calculation of $\pi (z),$
for which the discriminant of the square root in the last equation is set to
zero. This results in the polynomial $\pi (z)$ which is dependent on the
transformation function $z(r).$ Also, the parameter $\lambda $ defined in
Eq. (9) takes the form%
\begin{equation}
\lambda =\lambda _{n}=-n\tau ^{\prime }(z)-\frac{1}{2}n\left( n-1\right)
\sigma ^{\prime \prime }(z),\ \ \ n=0,1,2,\cdots .
\end{equation}%
At the end, the energy equation and consequently it's eigenvalues can be
obtained by comparing Eqs. (9) and (10).

Let us now construct a parametric generalization of the NU method valid for
any central and non-central exponential-type potential. Comparing the
following generalized hypergeometric equation
\begin{equation}
\left[ z\left( 1-c_{3}z\right) \right] ^{2}u^{\prime \prime }(z)+\left[
z\left( 1-c_{3}z\right) \left( c_{1}-c_{2}z\right) \right] u^{\prime
}(z)+\left( -B_{1}z^{2}+B_{2}z-B_{3}\right) u(z)=0,
\end{equation}%
with Eq. (1), we obtain%
\begin{equation}
\widetilde{\tau }(z)=c_{1}-c_{2}z,\text{ }\sigma (z)=z\left( 1-c_{3}z\right)
,\text{ }\widetilde{\sigma }(z)=-B_{1}z^{2}+B_{2}z-B_{3},
\end{equation}%
where the parameters $c_{j}$ and $B_{j}$ ($j=1,2,3$) are to be determined
during the solution procedure. Thus, by following the method, we may obtain
all the analytic polynomials and their relevant constants necessary for the
solution of a radial wave equation. These analytic expressions are cited in
Appendix A.

\section{Solutions of the Dirac-Generalized WS Problem}

The Dirac equation for fermionic massive spin-$1/2$ particles moving in an
attractive scalar potential $S(r)$ and a repulsive vector potential $V(r)$
can be written as [35]
\begin{equation}
\left[ c\mathbf{\alpha }\cdot \mathbf{p+\beta }\left( mc^{2}+S(r)\right)
+V(r)-E\right] \psi (\mathbf{r})=0,\text{ }\psi (\mathbf{r})=\psi (r,\theta
,\phi ),
\end{equation}%
where $E$ is the relativistic energy of the system, $\mathbf{p}=-i\hbar
\mathbf{\nabla }$ is the momentum operator, and $\mathbf{\alpha }$ and $%
\mathbf{\beta }$ are $4\times 4$ Dirac matrices%
\begin{equation}
\mathbf{\alpha =}\left(
\begin{array}{cc}
0 & \mathbf{\sigma } \\
\mathbf{\sigma } & 0%
\end{array}%
\right) ,\text{ }\mathbf{\beta =}\left(
\begin{array}{cc}
\mathbf{I} & 0 \\
0 & -\mathbf{I}%
\end{array}%
\right) ,\text{ }\sigma _{1}\mathbf{=}\left(
\begin{array}{cc}
0 & 1 \\
1 & 0%
\end{array}%
\right) ,\text{ }\sigma _{2}\mathbf{=}\left(
\begin{array}{cc}
0 & -i \\
i & 0%
\end{array}%
\right) ,\text{ }\sigma _{3}\mathbf{=}\left(
\begin{array}{cc}
1 & 0 \\
0 & -1%
\end{array}%
\right) .
\end{equation}%
where $\mathbf{I}$ denotes the $2\times 2$ identity matrix and $\mathbf{%
\sigma }$ are three-vector Pauli spin matrices. For a spherical symmetrical
nuclei, the total angular momentum operator of the nuclei $\mathbf{J}$ and
spin-orbit matrix operator $\mathbf{K}=-\mathbf{\beta }\left( \mathbf{\sigma
}\cdot \mathbf{L}+\mathbf{I}\right) $ commute with the Dirac Hamiltonian,
where $\mathbf{L}$ is the orbital angular momentum operator. The spinor
wavefunctions can be classified according to the radial quantum number $n$
and the spin-orbit quantum number $\kappa $ and can be written using the
Pauli-Dirac representation:%
\begin{equation}
\psi _{n\kappa }=\left(
\begin{array}{c}
f_{n\kappa } \\
g_{n\kappa }%
\end{array}%
\right) =\frac{1}{r}\left(
\begin{array}{c}
F_{n\kappa }(r)Y_{jm\kappa }^{l}(\widehat{r}) \\
iG_{n\kappa }(r)Y_{jm(-\kappa )}^{\widetilde{l}}(\widehat{r})%
\end{array}%
\right) ,
\end{equation}%
where $F_{n\kappa }(r)$ and $G_{n\kappa }(r)$ are the radial wave functions
of the upper- and lower-spinor components, respectively, $Y_{jm\kappa }^{l}(%
\widehat{r})$ and $Y_{jm(-\kappa )}^{\widetilde{l}}(\widehat{r})$ denote the
spin spherical harmonic functions coupled to the total angular momentum $j$
and it's projection $m$ on the $z$ axis and $l(l+1)=\kappa \left( \kappa
+1\right) $ and $\widetilde{l}(\widetilde{l}+1)=\kappa \left( \kappa
-1\right) $. The quantum number $\kappa $ is related to the quantum numbers
for spin symmetry $l$ and pseudospin symmetry $\widetilde{l}$ as%
\begin{equation}
\kappa =\left\{
\begin{array}{cccc}
-\left( l+1\right) =-\left( j+\frac{1}{2}\right) , & (s_{1/2},p_{3/2},\text{%
\textit{etc.}}), & \text{ }j=l+\frac{1}{2}, & \text{aligned spin }\left(
\kappa <0\right) , \\
+l=+\left( j+\frac{1}{2}\right) , & \text{ }(p_{1/2},d_{3/2},\text{\textit{%
etc.}}), & \text{ }j=l-\frac{1}{2}, & \text{unaligned spin }\left( \kappa
>0\right) .%
\end{array}%
\right.
\end{equation}%
and the quasi-degenerate doublet structure can be expressed in terms of a
pseudospin angular momentum $\widetilde{s}=1/2$ and pseudo-orbital angular
momentum $\widetilde{l}$ which is defined as
\begin{equation}
\kappa =\left\{
\begin{array}{cccc}
-\widetilde{l}=-\left( j+\frac{1}{2}\right) , & (s_{1/2},p_{3/2},\text{%
\textit{etc.}}), & j=\widetilde{l}-1/2, & \text{aligned spin }\left( \kappa
<0\right) , \\
+\left( \widetilde{l}+1\right) =+\left( j+\frac{1}{2}\right) , & \text{ }%
(p_{1/2},d_{3/2},\text{\textit{etc.}}), & \ j=\widetilde{l}+1/2, & \text{%
unaligned spin }\left( \kappa >0\right) .%
\end{array}%
\right.
\end{equation}%
where $\kappa =\pm 1,\pm 2,\cdots .$ For example, ($3s_{1/2},2d_{3/2}$) and (%
$3\widetilde{p}_{1/2},2\widetilde{p}_{3/2}$) can be considered as pseudospin
doublets.

Substituting Eq. (15) into Eq. (13), we obtain two radial coupled Dirac
equations for the spinor components
\begin{subequations}
\begin{equation}
\left( \frac{d}{dr}+\frac{\kappa }{r}\right) F_{n\kappa }(r)=\left[
mc^{2}+E_{n\kappa }-\Delta (r)\right] G_{n\kappa }(r),
\end{equation}%
\begin{equation}
\left( \frac{d}{dr}-\frac{\kappa }{r}\right) G_{n\kappa }(r)=\left[
mc^{2}-E_{n\kappa }+\Sigma (r)\right] F_{n\kappa }(r),
\end{equation}%
where $\Delta (r)=V(r)-S(r)$ and $\Sigma (r)=V(r)+S(r)$ are the difference
and sum potentials, respectively. By eliminating $G_{n\kappa }(r)$ in Eq.
(18a) and $F_{n\kappa }(r)$ in Eq. (18b), one is able to obtain two
second-order differential equations for the upper- and lower-spinor
components as follows:
\end{subequations}
\begin{subequations}
\begin{equation}
\left\{ \frac{d^{2}}{dr^{2}}-\frac{\kappa \left( \kappa +1\right) }{r^{2}}-%
\frac{1}{\hbar ^{2}c^{2}}\left[ U_{-}(r)U_{+}(r)-\frac{\frac{d\Delta (r)}{dr}%
\left( \frac{d}{dr}+\frac{\kappa }{r}\right) }{mc^{2}+E_{n\kappa }-\Delta (r)%
}\right] \right\} F_{n\kappa }(r)=0,\text{ }F_{n\kappa }(0)=0,
\end{equation}%
\begin{equation}
\left\{ \frac{d^{2}}{dr^{2}}-\frac{\kappa \left( \kappa -1\right) }{r^{2}}-%
\frac{1}{\hbar ^{2}c^{2}}\left[ U_{-}(r)U_{+}(r)+\frac{\frac{d\Sigma (r)}{dr}%
\left( \frac{d}{dr}-\frac{\kappa }{r}\right) }{mc^{2}-E_{n\kappa }+\Sigma (r)%
}\right] \right\} G_{n\kappa }(r)=0,\text{ }G_{n\kappa }(0)=0,
\end{equation}%
where $U_{-}(r)=mc^{2}+E_{n\kappa }-\Delta (r)$ and $U_{+}(r)=mc^{2}-E_{n%
\kappa }+\Sigma (r),$ are the difference and the sum functions,
respectively. From the above equations, the energy eigenvalues depend on the
quantum numbers $n$ and $\kappa $, and also the pseudo-orbital angular
quantum number $\widetilde{l}$ according to $\kappa (\kappa -1)=\widetilde{l}%
(\widetilde{l}+1),$ which implies that $j=\widetilde{l}\pm 1/2$ are
degenerate for $\widetilde{l}\neq 0.$ The above non-linear radial wave
equations having very complicated solutions are required to satisfy the
necessary boundary conditions $F_{n\kappa }(0)=G_{n\kappa }(0)$ and $%
F_{n\kappa }(\infty )=G_{n\kappa }(\infty )$ for bound state solutions$.$

In this context, we take the sum potential in the form of an attractive
generalized WS potential, i.e., $\Sigma (r)=V_{GWS}(r)$ [36]. The
interaction among nuclei is commonly described by using a potential which
consists of the Coulomb and the nuclear potentials. It is usually taken in
the form of WS potential. Here we take the following Hermitian real-valued
generalized WS potential which is specified by the shape (deformation)
parameter, $q,$ [13,22,37]
\end{subequations}
\begin{equation}
V_{GWS}(x)=V_{q}(x)=-V_{0}\frac{e^{-\alpha x}}{1+qe^{-\alpha x}},\text{ \ }%
R_{0}/a\equiv \alpha ,\text{ }\left( r-R_{0}\right) /R_{0}\equiv x,\text{ }%
R_{0}\gg a,\text{ }q>0
\end{equation}%
where $r\in (0,\infty )$ or $x\in (-1,\infty )$ refers to the center-of-mass
distance between the projectile and the target nuclei. The relevant
parameters of the inter-nuclear potential are given as follows: $R_{\text{0}%
}=r_{0}A_{0}^{1/3}$ is to define the confinement barrier position value of
the corresponding spherical nucleus or the range of the potential well, $%
A_{0}$ is the atomic mass number of target nucleus, $r_{0}$ is the radius
parameter, the parameter $V_{0}$ is the potential depth, $a$ is the surface
thickness and has to control it's slope, which is usually adjusted to the
experimental values of ionization energies. Note further, $q$ is a real
shape (deformation) parameter, the strength of the exponential part other
than unity, set to determine the shape of potential and is arbitrarily taken
to be a real constant within the potential. In addition, it should be noted
that the spatial coordinates in the potential are not deformed and thus the
potential still remains spherical.

It is worth noting that under radial coordinate transformation, $%
r\rightarrow r+\Delta ,$ then the generalized WS potential in Eq. (20)
changes into the standard WS potential (when $q$ is taken equal to $1$ in
the calculation$)$ but with the displacement parameter $\Delta $ satisfies
the expression $\exp (\Delta /a)=q$ and with a field strength $V_{0}^{\prime
}=V_{0}\exp (-\Delta /a)$ [13,22]. The sense of generalization or
deformation of the potential becomes clear. For completeness, it could be
stated that if $\Delta $ is positive (corresponding to $q>1)$ then one may
need to impose the condition on the choice of $\Delta ,$ that is, $%
\left\vert \Delta \right\vert \ll R_{0}.$

Obviously, for some specific $q$ values this potential reduces to the
well-known types, such as for $q=0$ to the exponential potential and for $%
q=-1$ and $a=\delta ^{-1}$ to the generalized Hulth\'{e}n potential (cf.
[11,13,22] and the references therein). Obviously, the solutions in Ref.
[22] are at best valid for $R_{0}=0,$ in which the potential can be expanded
in terms of hyperbolic functions [19,22]. The standard WS potential turns to
become the well-known Rosen-Morse potential shifted by the term \ $-V_{0}/2$
(cf. Ref. [22] and the references therein)$,$ that is, $V_{SRM}(r)=-V_{1}%
\sec h^{2}(\alpha x)+V_{2}\tanh (\alpha x)-V_{3},$ where $V_{1}=C/4$ and $%
V_{2}=V_{3}=V_{0}/2$ [32].

\subsection{Spin symmetric solution}

In the case of exact spin symmetry $S(r)\sim V(r)$ ($d\Delta (r)/dr=0,$
\textit{i.e}., $\Delta (r)=A=$ constant), Eq. (19a) can be approximately
written as
\begin{equation}
\left\{ \frac{d^{2}}{dr^{2}}-\frac{\kappa \left( \kappa +1\right) }{r^{2}}-%
\frac{1}{\hbar ^{2}c^{2}}\left[ m(r)c^{2}+E_{n\kappa }-A\right] \left[
m(r)c^{2}-E_{n\kappa }+\Sigma (r)\right] \right\} F_{n\kappa }(r)=0,
\end{equation}%
where $\kappa \left( \kappa +1\right) =l\left( l+1\right) ,$ $\kappa =l$ for
$\kappa <0$ and $\kappa =-\left( l+1\right) $ for $\kappa >0.$ The spin
symmetric energy eigenvalues depend on $n$ and $\kappa ,$ \textit{i.e.}, $%
E_{n\kappa }=E(n,\kappa \left( \kappa +1\right) ).$ For $l\neq 0,$ the
states with $j=l\pm 1/2$ are degenerate. This is the exact spin symmetry.
Taking the $\Sigma (r)=2V(r)\rightarrow V_{GWS}(r)$ as mentioned in Ref.
[36] enables one to reduce the resulting relativistic solutions into their
non-relativistic limit under appropriate transformations. We are set out to
obtain bound state solutions (relativistic energy spectrum and upper- and
lower spinor wavefunctions) of a spin-zero particle for a four parameter \{$%
V_{0},q,a,R_{0}$\} generalized WS potential by means of the NU method.
Moreover, if $\kappa $ is not too large, the case of the vibrations of small
amplitude about the minimum, we can then use the approximate expansion of
the centrifugal potential near the minimum point $r=R_{0}$ $(x=0)$ as [38]%
\begin{equation*}
\frac{\kappa \left( \kappa +1\right) }{r^{2}}=\frac{\kappa \left( \kappa
+1\right) }{R_{0}^{2}}\left( 1+x\right) ^{-2}
\end{equation*}%
\begin{equation}
\approx \frac{\kappa \left( \kappa +1\right) }{R_{0}^{2}}\left\{ D_{0}+D_{1}%
\frac{-\exp (-\alpha x)}{1+q\exp (-\alpha x)}+D_{2}\left[ \frac{-\exp
(-\alpha x)}{1+q\exp (-\alpha x)}\right] ^{2}\right\} ,
\end{equation}%
where
\begin{subequations}
\begin{equation}
D_{0}=1-\left[ \frac{1+\exp (-\alpha R_{0})}{\alpha R_{0}}\right] ^{2}\left[
\frac{4\alpha R_{0}}{1+\exp (-\alpha R_{0})}-3-\alpha R_{0}\right] ,
\end{equation}%
\begin{equation}
D_{1}=2\left[ \exp (\alpha R_{0})+1\right] \left\{ 3\left[ \frac{1+\exp
(-\alpha R_{0})}{\alpha R_{0}}\right] -\left( 3+\alpha R_{0}\right) \left[
\frac{1+\exp (-\alpha R_{0})}{\alpha R_{0}}\right] \right\} ,
\end{equation}%
\begin{equation}
D_{2}=\left[ \exp (\alpha R_{0})+1\right] ^{2}\left[ \frac{1+\exp (-\alpha
R_{0})}{\alpha R_{0}}\right] ^{2}\left[ 3+\alpha R_{0}-\frac{2\alpha R_{0}}{%
1+\exp (-\alpha R_{0})}\right] ,
\end{equation}%
and higher order terms are neglected. It is worth noting that for $\kappa
\neq 0$ case$,$ we have to use an approximation for the centrifugal term
similar to the non-relativistic cases which is valid only for $q=1$ value
[6,38]. However, for $s$-waves, we remark that the problem can be solved
exactly and the solution is valid for any deformation parameter $q.$

We define the following new dimensionless parameter, z$(x)=-e^{-\alpha x}\in
\lbrack -e^{R_{0}/a},0]$, which maintains the finiteness of the transformed
wave functions on the boundary conditions$.$ Thus$,$ u$\sin $g Eqs. (22) and
(23) , we can reduce Eq. (21) to generalized equation of the hypergeometric
type for the upper-spinor component $F_{n\kappa }(r),$%
\end{subequations}
\begin{equation}
\left[ \frac{d^{2}}{dz^{2}}+\frac{(1-qz)}{z(1-qz)}\frac{d}{dz}+\frac{1}{%
\left[ z(1-qz)\right] ^{2}}\left( -\beta _{1}z^{2}+\beta _{2}z-\varepsilon
_{n\kappa }^{2}\right) \right] F_{n\kappa }(z)=0,\text{ }F_{n\kappa
}(0)=F_{n\kappa }(\infty )=0,
\end{equation}%
where $F_{n\kappa }(r)=F_{n\kappa }(z)$ and we introduce the definitions
\begin{subequations}
\begin{equation}
\varepsilon _{n\kappa }=\sqrt{\frac{a^{2}}{\hbar ^{2}c^{2}}\left[
m^{2}c^{4}-E_{n\kappa }^{2}-A\left( mc^{2}-E_{n\kappa }\right) \right] +%
\frac{\omega a^{2}}{r_{0}^{2}}D_{0}}>0,\text{ }
\end{equation}%
\begin{equation}
\beta _{1}=q^{2}\varepsilon _{n\kappa }^{2}-\frac{qa^{2}V_{0}}{\hbar
^{2}c^{2}}\left( mc^{2}+E_{n\kappa }-A\right) -\frac{\omega a^{2}}{r_{0}^{2}}%
\left( qD_{1}-D_{2}\right) ,\text{ }
\end{equation}%
\begin{equation}
\beta _{2}=2q\varepsilon _{n\kappa }^{2}-\frac{a^{2}V_{0}}{\hbar ^{2}c^{2}}%
\left( mc^{2}+E_{n\kappa }-A\right) -\frac{\omega a^{2}}{r_{0}^{2}}D_{1},%
\text{ }
\end{equation}%
with $\omega =\kappa (\kappa +1)$ where $\kappa =\pm 1,\pm 2,\cdots ,$ for
bound states (i.e., real $\varepsilon _{n\kappa }$). Before we can proceed,
it is necessary to compare the last equation with Eq. (1) to obtain the
following polynomials:
\end{subequations}
\begin{equation}
\widetilde{\tau }(z)=1-qz,\text{ \ \ \ }\sigma (z)=z(1-qz),\text{ \ \ }%
\widetilde{\sigma }(z)=-\beta _{1}z^{2}+\beta _{2}z-\varepsilon _{n\kappa
}^{2}.
\end{equation}%
We follow Appendix A to calculate the specific values of the parametric
constants and then display them in Table 1 for the present potential model.
Also, with the aid of Table 1, the key polynomials given in Appendix A take
the following particular analytic forms:
\begin{equation}
\pi (z)=\varepsilon _{n\kappa }-\frac{q}{2}\left( 1+2\varepsilon _{n\kappa
}+\xi \right) z,
\end{equation}%
\begin{equation}
k=\beta _{2}-q\left( 2\varepsilon _{n\kappa }+\xi \right) \varepsilon
_{n\kappa },
\end{equation}%
\begin{equation}
\tau (z)=1+2\varepsilon _{n\kappa }-q\left( 2+2\varepsilon _{n\kappa }+\xi
\right) z,
\end{equation}%
where $\tau ^{\prime }(z)=-q\left( 2+2\varepsilon _{n\kappa }+\xi \right) <0$
with $\xi =\sqrt{1+\frac{4\omega a^{2}}{q^{2}R_{0}^{2}}D_{2}}.$ We insert
the values of the constants given in Table 1 into the energy equation cited
in Appendix A and then obtain%
\begin{equation}
\varepsilon _{n\kappa }=-\left( \frac{a}{q}\right) ^{2}\left[ \frac{\frac{%
qV_{0}}{\hbar ^{2}c^{2}}\left( mc^{2}+E_{n\kappa }-A\right) +\frac{\omega }{%
R_{0}^{2}}\left( qD_{1}-D_{2}\right) }{\left( 1+2n+\sqrt{1+\frac{4\omega
a^{2}}{q^{2}R_{0}^{2}}D_{2}}\right) }+\left( \frac{q}{2a}\right) ^{2}\left(
1+2n+\sqrt{1+\frac{4\omega a^{2}}{q^{2}R_{0}^{2}}D_{2}}\right) \right] ,
\end{equation}%
where $\kappa =\pm 1,\pm 2,\cdots .$ Hence, the above equation gives
explicitly the energy equation with exact spin symmetry for arbitrary
spin-orbit coupling quantum $\kappa $ of the Dirac equation as follows%
\begin{equation*}
\left[ m^{2}c^{4}-E_{n\kappa }^{2}-A\left( mc^{2}-E_{n\kappa }\right) \right]
=-\frac{\hbar ^{2}c^{2}\omega }{R_{0}^{2}}D_{0}
\end{equation*}%
\begin{equation}
+\left( \frac{a\hbar c}{q}\right) ^{2}\left[ \frac{\frac{V_{0}}{\hbar
^{2}c^{2}}\left( mc^{2}+E_{n\kappa }-A\right) +\frac{\omega }{qR_{0}^{2}}%
\left( qD_{1}-D_{2}\right) }{\left( 1+2n+\sqrt{1+\frac{4\omega a^{2}}{%
q^{2}R_{0}^{2}}D_{2}}\right) }+\frac{q}{4a^{2}}\left( 1+2n+\sqrt{1+\frac{%
4\omega a^{2}}{q^{2}R_{0}^{2}}D_{2}}\right) \right] ^{2},
\end{equation}%
The energy level $E_{n\kappa }$ is determined by energy equation (31), which
is a rather complicated transcendental equation. Now, let us consider a few
special cases of much concern. (i) If we choose $q=1,$ the potential (20)
turns to the shifted WS potential:%
\begin{equation}
V(x)=-V_{0}+\frac{V_{0}}{1+e^{-\alpha x}},
\end{equation}%
and then it's energy spectra yield%
\begin{equation*}
\left( mc^{2}-E_{n\kappa }\right) \left( mc^{2}+E_{n\kappa }-A\right) =-%
\frac{\hbar ^{2}c^{2}\omega }{R_{0}^{2}}D_{0}
\end{equation*}%
\begin{equation}
+\hbar ^{2}c^{2}a^{2}\left[ \frac{\frac{V_{0}}{\hbar ^{2}c^{2}}\left(
mc^{2}+E_{n\kappa }-A\right) +\frac{\omega }{R_{0}^{2}}\left(
D_{1}-D_{2}\right) }{\left( 1+2n+\sqrt{1+\frac{4\omega a^{2}}{R_{0}^{2}}D_{2}%
}\right) }+\frac{1}{4a^{2}}\left( 1+2n+\sqrt{1+\frac{4\omega a^{2}}{R_{0}^{2}%
}D_{2}}\right) \right] ^{2}.
\end{equation}%
(ii) If we choose $q=-1,$ the potential (20) turns to the standard shifted
Hulth\'{e}n potential:%
\begin{equation}
V(x)=V_{0}-\frac{V_{0}}{1-e^{-\alpha x}},
\end{equation}%
and then the resulting energy eigenvalues become%
\begin{equation*}
\left( mc^{2}-E_{n\kappa }\right) \left( mc^{2}+E_{n\kappa }-A\right) =-%
\frac{\hbar ^{2}c^{2}\omega }{R_{0}^{2}}D_{0}
\end{equation*}%
\begin{equation}
+\hbar ^{2}c^{2}a^{2}\left[ \frac{\frac{V_{0}}{\hbar ^{2}c^{2}}\left(
mc^{2}+E_{n\kappa }-A\right) +\frac{\omega }{R_{0}^{2}}\left(
D_{1}+D_{2}\right) }{\left( 1+2n+\sqrt{1+\frac{4\omega a^{2}}{R_{0}^{2}}D_{2}%
}\right) }-\frac{1}{4a^{2}}\left( 1+2n+\sqrt{1+\frac{4\omega a^{2}}{R_{0}^{2}%
}D_{2}}\right) \right] ^{2}.
\end{equation}%
(iii) If we choose $q\rightarrow 0,$ the potential (20) turns to the
exponential potential:%
\begin{equation}
V(x)=-V_{0}e^{-\alpha x},
\end{equation}%
the eigenvalues expression (31) does not give an explicit form, i.e., the NU
method is not applicable to the exponential potential (36). Note that for
this potential there is no explicit form of the energy expression of bound
states for Schr\"{o}dinger [9], KG [12] and also Dirac [8] equations.

In addition, for the $s$-wave ($\kappa =-1$) and $V(r)=S(r)$ (i.e., $A=0)$,
we obtain%
\begin{equation}
m^{2}c^{4}-E_{n(-1)}^{2}=\frac{\hbar ^{2}c^{2}a^{2}}{q^{2}}\left[ \frac{V_{0}%
}{2\hbar ^{2}c^{2}}\frac{\left( mc^{2}+E_{n(-1)}\right) }{n+1}+\frac{q(n+1)}{%
2a^{2}}\right] ^{2},\text{ }n=0,1,2,\cdots ,
\end{equation}%
and it can be seen easily that while the field strength $V_{0}\rightarrow 0,$
the energy states yield:%
\begin{equation}
E_{n(-1)}^{\pm }=\pm \frac{1}{2a}\sqrt{4a^{2}m^{2}c^{4}-\hbar
^{2}c^{2}\left( n+1\right) ^{2}},\text{ }n=0,1,2,\cdots ,
\end{equation}%
for particles and anti-particles. Note that in the above equation there
exist bound states for the ground and excited states $(n=0,1)$ which are $%
E_{0}=\pm \sqrt{3}mc^{2}/2$ and $E_{1}=0,$ respectively, for positive $q$
values and $a=\lambda _{c},$ where $\lambda _{c}=\hbar /mc$ denotes the
Compton wavelength of the Dirac particle. Otherwise, there are no bound
states for $n\geq 2$ states.

On the other hand, for the same value of $\alpha $ and negative $q$ values
when $V_{0}\rightarrow 0,$ all energy eigenvalues go to zero. If the value
of $q$ is increasing, all positive bound states go to zero, from (38),
asymptotically.

An inspection of the energy expression given by Eq. (37), for any given $%
\alpha ,$ shows that we deal with a family of generalized WS potentials. The
sign of $V_{0}$ does not effect the bound states. The spectrum consists of
complex eigenvalues depending on $q.$ As we shall see the role played by the
range parameter $\alpha $ is very crucial in this regard. Of course, it is
clear that by imposing appropriate changes in the parameters $\left\{ \alpha
,V_{0},q\right\} ,$ the energy spectrum in Eq. (37) for any modified
parameter can be also calculated by resolving Dirac equation for every
parameter change.

The upper-spinor wave functions for $F_{n\kappa }(r)$ will be presented. In
order to establish $F_{n\kappa }(r),$ use will be made of Appendix A and
Table 1. Hence, the first part of wave functions reads:%
\begin{equation}
\phi _{n}(z)=z^{\varepsilon _{n\kappa }}(1-qz)^{\frac{1}{2}\left( 1+\xi
\right) },\text{ }\varepsilon _{n\kappa }>0,\text{ }\xi >-1.
\end{equation}%
In addition, to find the function, $y_{n}(z),$ which is the polynomial
solution of hypergeometric-type equation, we firstly calculate the weight
function:%
\begin{equation}
\rho (z)=z^{2\varepsilon _{n\kappa }}(1-qz)^{\xi }.
\end{equation}%
and thus the second part of wave functions (7) can be obtained as%
\begin{equation}
y_{n}(z)=D_{n}z^{-2\varepsilon _{n\kappa }}(1-qz)^{-\xi }\frac{d^{n}}{dz^{n}}%
\left[ z^{n+2\varepsilon _{n\kappa }}\left( 1-z\right) ^{n+\xi }\right] ,%
\text{ }\xi >-1,
\end{equation}%
where $D_{n}$ is a normalization constant$.$ In the limit $q\rightarrow 1,$
the polynomial solutions of $\ y_{n}(z)$ are expressed in terms of Jacobi
Polynomials, which is one of the classical orthogonal polynomials, with
weight function given by Eq. (40) for $z\in $ [$0,1],$ giving $%
y_{n}(z)\simeq P_{n}^{(2\varepsilon _{n\kappa },\xi )}(1-2z),$ $2\varepsilon
_{n\kappa },$ $\xi >-1$ [39]. Thus the associated uppercomponent $F_{n\kappa
}(z)$ for arbitrary the spin-orbit coupling quantum number $\kappa $ can be
obtained by substituting Eqs. (39) and (41) into Eq. (2) as%
\begin{equation*}
F_{n\kappa }(z)=\mathcal{N}_{n\kappa }z^{\varepsilon _{n\kappa }}(1-qz)^{%
\frac{1}{2}\left( 1+\xi \right) }P_{n}^{(2\varepsilon _{n\kappa },\xi
)}(1-2qz)
\end{equation*}%
\begin{equation}
=\mathcal{N}_{n\kappa }z^{\varepsilon _{n\kappa }}(1-qz)^{\frac{1}{2}\left(
1+\xi \right) }%
\begin{array}{c}
_{2}F_{1}%
\end{array}%
\left( -n,1+2\varepsilon _{n\kappa }+\xi +n;1+2\varepsilon _{n\kappa
};qz\right) ,
\end{equation}%
where $z(r)=-e^{-(r-R_{0})/a}$ and $\mathcal{N}_{n\kappa }$ are
normalization constants calculated in Appendix B.

Before presenting the corresponding lower-component $G_{n\kappa }(r),$ let
us recall a recurrence relation of hypergeometric function, which is used to
solve Eq. (18a) and present the corresponding lower component $G_{n\kappa
}(r),$%
\begin{equation}
\frac{d}{dz}\left[
\begin{array}{c}
_{2}F_{1}%
\end{array}%
\left( a;b;c;z\right) \right] =\left( \frac{ab}{c}\right)
\begin{array}{c}
_{2}F_{1}%
\end{array}%
\left( a+1;b+1;c+1;z\right) ,
\end{equation}%
with which the corresponding lower component $G_{n\kappa }(r)$ can be given
by solving Eq. (18a) as follows%
\begin{equation*}
G_{n\kappa }(r)=\frac{1}{mc^{2}+E_{n\kappa }-A}\left[ \frac{dF_{n\kappa }(r)%
}{dr}+\frac{\kappa }{r}F_{n\kappa }(r)\right]
\end{equation*}%
\begin{equation*}
=\frac{\mathcal{N}_{n\kappa }\left( -e^{-(r-R_{0})/a}\right) ^{\varepsilon
_{n\kappa }}(1+qe^{-(r-R_{0})/a})^{\frac{1}{2}\left( 1+\xi \right) }}{%
mc^{2}+E_{n\kappa }-A}\left[ \frac{\varepsilon _{n\kappa }}{a}-\frac{q\left(
1+\xi \right) e^{-(r-R_{0})/a}}{2a\left( 1+qe^{-(r-R_{0})/a}\right) }+\frac{%
\kappa }{r}\right]
\end{equation*}%
\begin{equation*}
\times
\begin{array}{c}
_{2}F_{1}%
\end{array}%
\left( -n,1+2\varepsilon _{n\kappa }+\xi +n;2\varepsilon _{n\kappa
}+1;-qe^{-(r-R_{0})/a}\right)
\end{equation*}%
\begin{equation*}
+\mathcal{N}_{n\kappa }\left[ \frac{qn\left( 1+2\varepsilon _{n\kappa }+\xi
+n\right) \left( -e^{-(r-R_{0})/a}\right) ^{\varepsilon _{n\kappa }+1}\left(
1+qe^{-(r-R_{0})/a}\right) ^{\frac{1}{2}\left( 1+\xi \right) }}{a\left(
2\varepsilon _{n\kappa }+1\right) \left( mc^{2}+E_{n\kappa }-A\right) }%
\right]
\end{equation*}%
\begin{equation}
\times
\begin{array}{c}
_{2}F_{1}%
\end{array}%
\left( -n+1;n+\xi +2\left( 1+\varepsilon _{n\kappa }\right) ;2\left(
1+\varepsilon _{n\kappa }\right) ;-qe^{-(r-R_{0})/a}\right) .
\end{equation}%
Here, it should be noted that the hypergeometric series $%
\begin{array}{c}
_{2}F_{1}%
\end{array}%
\left( -n,1+2\varepsilon _{n\kappa }+\xi +n;2\varepsilon _{n\kappa
}+1;-qe^{-(r-R_{0})/a}\right) $ does not terminate for $n=0$ and thus does
not diverge for all values of real parameters $\xi $ and $\varepsilon
_{n\kappa }.$

For $A>mc^{2}+E_{n\kappa }$ and $E_{n\kappa }<mc^{2}$ or $%
A<mc^{2}+E_{n\kappa }$ and $E_{n\kappa }>mc^{2},$ we note that parameters
given in Eq. (25a) turn to be imaginary, i.e., $\varepsilon _{n\kappa
}^{2}<0 $ in the $s$-state ($\kappa =-1$)$.$ As a result, the condition of
existing bound states are $\varepsilon _{n\kappa }>0$ and $\xi >0,$ that is
to say, in the case of $A>mc^{2}+E_{n\kappa }$ and $E_{n\kappa }<mc^{2},$
bound-states do not exist for some quantum number $\kappa $ such as the $s$%
-state ($\kappa =-1$)$.$ Of course, if these conditions are satisfied for
existing bound-states, the energy equation and wave functions are the same
as these given in Eq. (31) and Eqs. (42)-(44).

\subsection{Pseudospin symmetric solution}

Under the condition of the pseudospin symmetry $S(r)\sim -V(r)$ (\textit{i.e}%
., $d\Sigma (r)/dr=0,$ or $\Sigma (r)=A^{\prime }=$ constant), Eq. (19b) can
be exactly written as
\begin{equation}
\left\{ \frac{d^{2}}{dr^{2}}-\frac{\widetilde{\omega }}{r^{2}}-\frac{1}{%
\hbar ^{2}c^{2}}\left[ mc^{2}+E_{n\kappa }-\Delta (r)\right] \left[
mc^{2}-E_{n\kappa }+A^{\prime }\right] \right\} G_{n\kappa }(r)=0,
\end{equation}%
where $\widetilde{\omega }=\kappa \left( \kappa -1\right) =\widetilde{l}(%
\widetilde{l}+1),$ the energy eigenvalues $E_{n\kappa }$ depend only on $n$
and $\widetilde{l},$ \textit{i.e.}, $E_{n\kappa }=E(n,\widetilde{l}(%
\widetilde{l}+1)).$ Taking the $\Delta (r)=V_{GWS}(r)$ allows us to reduce
our results to the non-relativistic limit. For $\widetilde{l}\neq 0,$ the
states with $j=\widetilde{l}\pm 1/2$ are degenerate. This is the exact
pseudospin symmetry. We follow the procedures in the previous subsection to
obtain Dirac equation satisfying $G_{n\kappa }(r),$%
\begin{equation}
\left[ \frac{d^{2}}{dz^{2}}+\frac{(1-qz)}{z(1-qz)}\frac{d}{dz}+\frac{1}{%
\left[ z(1-qz)\right] ^{2}}\left( -\widetilde{\beta }_{1}z^{2}+\widetilde{%
\beta }_{2}z-\widetilde{\varepsilon }_{n\kappa }^{2}\right) \right]
G_{n\kappa }(z)=0,\text{ }G_{n\kappa }(0)=G_{n\kappa }(\infty )=0,
\end{equation}%
where $G_{n\kappa }(r)=G_{n\kappa }(z)$ and we have used the definitions
\begin{subequations}
\begin{equation}
\widetilde{\varepsilon }_{n\kappa }=\sqrt{\frac{a^{2}}{\hbar ^{2}c^{2}}\left[
m^{2}c^{4}-E_{n\kappa }^{2}+A^{\prime }\left( mc^{2}+E_{n\kappa }\right) %
\right] +\frac{\widetilde{\omega }a^{2}}{R_{0}^{2}}D_{0}}>0,\text{ }\kappa
=\pm 1,\pm 2,\cdots ,
\end{equation}%
\begin{equation}
\widetilde{\beta }_{1}=q^{2}\varepsilon _{n\kappa }^{2}+\frac{qa^{2}V_{0}}{%
\hbar ^{2}c^{2}}\left( mc^{2}-E_{n\kappa }+A^{\prime }\right) -\frac{%
\widetilde{\omega }a^{2}}{R_{0}^{2}}\left( qD_{1}-D_{2}\right) ,\text{ }
\end{equation}%
\begin{equation}
\widetilde{\beta }_{2}=2q\varepsilon _{n\kappa }^{2}+\frac{a^{2}V_{0}}{\hbar
^{2}c^{2}}\left( mc^{2}-E_{n\kappa }+A^{\prime }\right) -\frac{\widetilde{%
\omega }a^{2}}{R_{0}^{2}}D_{1}.
\end{equation}%
To avoid repetition in the solution of Eq. (46), a first inspection for the
relationship between the present set of parameters $(\widetilde{\varepsilon }%
_{n\kappa },\widetilde{\beta }_{1},\widetilde{\beta }_{2})$ and the previous
set $(\varepsilon _{n\kappa },\beta _{1},\beta _{2})$ tells us that the
negative energy solution for pseudospin symmetry, where $S(r)\sim -V(r),$
can be obtained directly from those of the positive energy solution above
for spin symmetry using the following parameter mapping [39-41]:
\end{subequations}
\begin{equation}
F_{n\kappa }(r)\leftrightarrow G_{n\kappa }(r),V(r)\rightarrow -V(r)\text{
(or }V_{0}\rightarrow -V_{0}\text{)},\text{ }E_{n\kappa }\rightarrow
-E_{n\kappa }\text{ and }A\rightarrow -A^{\prime }.
\end{equation}%
Following the previous results with the above transformations, we finally
arrive at the energy equation. The relativistic transcendental energy
equation is%
\begin{equation*}
\left[ m^{2}c^{4}-E_{n\kappa }^{2}+A^{\prime }\left( mc^{2}+E_{n\kappa
}\right) \right] =-\frac{\hbar ^{2}c^{2}\widetilde{\omega }}{R_{0}^{2}}D_{0}
\end{equation*}%
\begin{equation}
+\left( \frac{a\hbar c}{q}\right) ^{2}\left[ \frac{-\frac{V_{0}}{\hbar
^{2}c^{2}}\left( mc^{2}-E_{n\kappa }+A^{\prime }\right) +\frac{\widetilde{%
\omega }}{qr_{0}^{2}}\left( qD_{1}-D_{2}\right) }{\left( 1+2n+\sqrt{1+\frac{4%
\widetilde{\omega }a^{2}}{q^{2}R_{0}^{2}}D_{2}}\right) }+\frac{q}{4a^{2}}%
\left( 1+2n+\sqrt{1+\frac{4\widetilde{\omega }a^{2}}{q^{2}R_{0}^{2}}D_{2}}%
\right) \right] ^{2},
\end{equation}%
and the lower-spinor wave functions%
\begin{equation*}
G_{n\kappa }(z)=\widetilde{\mathcal{N}}_{n\kappa }z^{\widetilde{\varepsilon }%
_{n\kappa }}(1-qz)^{\frac{1}{2}\left( 1+\widetilde{\xi }\right) }P_{n}^{(2%
\widetilde{\varepsilon }_{n\kappa },\widetilde{\xi })}(1-2qz)
\end{equation*}%
\begin{equation}
=\widetilde{\mathcal{N}}_{n\kappa }z^{\widetilde{\varepsilon }_{n\kappa
}}(1-qz)^{\frac{1}{2}\left( 1+\widetilde{\xi }\right) }%
\begin{array}{c}
_{2}F_{1}%
\end{array}%
\left( -n,1+2\widetilde{\varepsilon }_{n\kappa }+\widetilde{\xi }+n;2%
\widetilde{\varepsilon }_{n\kappa }+1;qz\right) ,
\end{equation}%
with $\widetilde{\xi }=\sqrt{1+\frac{4\widetilde{\omega }a^{2}}{%
q^{2}R_{0}^{2}}D_{2}}$ and $\widetilde{\varepsilon }_{n\kappa }$ is defined
in Eq. (47a).

\section{Discussions}

Now, let us study three special cases. We first study the $s(\widetilde{s})$%
-states ($l=\widetilde{l}=0,$ i.e., $\kappa =\mp 1$ ). In this case, we have
the spin-orbit coupling term $\kappa (\kappa +1)/r^{2}=0,$ and also the
corresponding approximation to it in Eq. (22). The corresponding energy
equation reduces to the $s$-states ($\kappa =-1$), i.e.,%
\begin{equation}
\left[ m^{2}c^{4}-E_{n(-1)}^{2}-A\left( mc^{2}-E_{n(-1)}\right) \right] =%
\frac{\hbar ^{2}c^{2}}{4}\left[ \frac{aV_{0}\left( mc^{2}+E_{n(-1)}-A\right)
}{q\hbar ^{2}c^{2}\left( 1+n\right) }+\frac{1}{a}\left( 1+n\right) \right]
^{2},
\end{equation}%
and the upper-spinor component of the wave functions:%
\begin{equation*}
F_{n}(z)=\mathcal{N}_{n}z^{\varepsilon _{n}}(1-qz)P_{n}^{(2\varepsilon
_{n},1)}(1-2qz)
\end{equation*}%
\begin{equation}
=\mathcal{N}_{n}z^{\varepsilon _{n}}(1-qz)%
\begin{array}{c}
_{2}F_{1}%
\end{array}%
\left( -n,2(1+\varepsilon _{n})+n;1+2\varepsilon _{n};qz\right) ,
\end{equation}%
with%
\begin{equation}
\varepsilon _{n}=\frac{a}{\hbar c}\sqrt{m^{2}c^{4}-E_{n(-1)}^{2}-A\left(
mc^{2}-E_{n(-1)}\right) }>0,\text{ }
\end{equation}%
where $\mathcal{N}_{n}$ is calculated in Appendix B. As mentioned above, in
the $s$-wave ($\kappa =-1$) the condition of existing bound-states is for $%
A<mc^{2}+E_{n(-1)}$ and $E_{n(-1)}<mc^{2}.$ Furthermore, in the
nonrelativistic limit with the mapping $mc^{2}-E_{n(-1)}\rightarrow -E_{n0}$
and $\frac{1}{\hbar ^{2}c^{2}}\left( mc^{2}+E_{n(-1)}\right) \rightarrow
\frac{2\mu }{\hbar ^{2}},$ then we have%
\begin{equation}
E_{n}=-\frac{\hbar ^{2}a^{2}}{8\mu }\left[ \frac{2\mu }{\hbar ^{2}}\frac{%
V_{0}}{q\left( 1+n\right) }+\frac{\left( 1+n\right) }{a^{2}}\right] ^{2},
\end{equation}%
Second, we study the special case $\Delta (r)=A=0.$ If so we have $%
V(r)=S(r)=1/2\Sigma (r)$ and it turns to the KG solution. Obviously, in this
case the energy equation given in Eq. (31) reduces to the energy equation of
arbitrary $\kappa $ state Dirac equation for equal scalar and vector WS
potential as follows%
\begin{equation*}
m^{2}c^{4}-E_{n\kappa }^{2}=-\frac{\hbar ^{2}c^{2}\omega }{R_{0}^{2}}D_{0}
\end{equation*}%
\begin{equation}
+\left( \frac{a\hbar c}{q}\right) ^{2}\left[ \frac{\frac{V_{0}}{\hbar
^{2}c^{2}}\left( mc^{2}+E_{n\kappa }\right) +\frac{\omega }{qR_{0}^{2}}%
\left( qD_{1}-D_{2}\right) }{\left( 1+2n+\sqrt{1+\frac{4\omega a^{2}}{%
q^{2}R_{0}^{2}}D_{2}}\right) }+\frac{q}{4a^{2}}\left( 1+2n+\sqrt{1+\frac{%
4\omega a^{2}}{q^{2}R_{0}^{2}}D_{2}}\right) \right] ^{2},
\end{equation}%
and the upper component of the wave functions%
\begin{equation*}
F_{n\kappa }(z)=\mathcal{N}_{n\kappa }z^{\varepsilon _{nl}}(1-qz)^{\frac{1}{2%
}\left( 1+\xi _{0}\right) }P_{n}^{(2\varepsilon _{nl},\xi _{0})}(1-2qz)
\end{equation*}%
\begin{equation}
=\mathcal{N}_{n\kappa }z^{\varepsilon _{nl}}(1-qz)^{\frac{1}{2}\left( 1+\xi
_{0}\right) }%
\begin{array}{c}
_{2}F_{1}%
\end{array}%
\left( -n,1+2\varepsilon _{n\kappa }+\xi _{0}+n;2\varepsilon _{n\kappa
}+1;qz\right) ,
\end{equation}%
\begin{equation}
\varepsilon _{n\kappa }=\sqrt{\frac{a^{2}}{\hbar ^{2}c^{2}}\left(
m^{2}c^{4}-E_{n\kappa }^{2}\right) +\frac{\omega a^{2}}{R_{0}^{2}}D_{0}},%
\text{ }\xi _{0}=\sqrt{1+\frac{4\kappa (\kappa +1)a^{2}}{q^{2}R_{0}^{2}}D_{2}%
}
\end{equation}%
where $E_{n\kappa }^{2}\leq m^{2}c^{4}+\frac{\omega \hbar ^{2}c^{2}}{%
R_{0}^{2}}D_{0}$ is the essential condition for existing bound-states.

Third, the non-relativistic energy state limit for arbitrary $l$ state are%
\begin{equation*}
E_{nl}=\frac{\hbar ^{2}l(l+1)}{2\mu R_{0}^{2}}D_{0}
\end{equation*}%
\begin{equation}
-\frac{\hbar ^{2}a^{2}}{2\mu }\left[ \frac{\frac{2\mu }{\hbar ^{2}}\frac{%
V_{0}}{q}+\frac{l(l+1)}{q^{2}R_{0}^{2}}\left( qD_{1}-D_{2}\right) }{\left(
1+2n+\sqrt{1+\frac{4l(l+1)a^{2}}{q^{2}R_{0}^{2}}D_{2}}\right) }+\frac{1}{%
4a^{2}}\left( 1+2n+\sqrt{1+\frac{4l(l+1)a^{2}}{q^{2}R_{0}^{2}}D_{2}}\right) %
\right] ^{2}.
\end{equation}%
The above result is identical to Eq. (23) in Ref. [22] where Berkdemir
\textit{et al.} used the usual approximation to the centrifugal term in the
potential expression (10) (cf. [42] and Eq. (2) in J. Math. Chem. 42, 461
(2007)).

It is worthy to note that in the calculations of Ref. [22], $R_{0}$ was
neglected. Accordingly, the solutions of the energy spectra Eq. (23) of the
original paper [22] are at best valid for $R_{0}=0$ in which case the
standard WS potential Eq. (10) in the original paper reduces to the shifted
Rosen-Morse (RM) potential (cf. Eq. (3) in Phys. Rev. C 74, 039902(E)
(2006)). The additional potential besides the standard Ws potential
considered by Berkdemir \textit{et al.} [22] provides the flexibility to
construct the surface structure of the related nucleus [42]. Thus, the
non-relativistic solutions obtained in [22] are only reasonable for the
hyperbolic [43] exponential (RM) potential [32], not WS potential. This
clear when we rewrite Eq. (20) in the following form
\begin{equation}
V_{GWS}(r)=-V_{0}^{\prime }\frac{e^{-\alpha r}}{1+qe^{-\alpha r}},\text{ }%
\alpha =\frac{1}{a},\text{ }V_{0}^{\prime }=V_{0}e^{R_{0}/a},\text{ }%
q=e^{R_{0}/a},\text{ }r\in (0,\infty ),
\end{equation}%
when $R_{0}=0,$ it implies that $q=1$ and then the above potential reduces
to the standard WS-type potential. In addition, the authors of Ref. [22]
approximated the centrifugal potential term $\frac{l(l+1)}{r^{2}}\approx
l(l+1)\alpha ^{2}\frac{e^{-r\alpha r}}{\left( 1-e^{-r\alpha r}\right) ^{2}}$
[13,43]$,$ where $C=l(l+1)\alpha ^{2}$ [20]. However, in the present work,
Eq. (58) contains the width of the potential $R_{0}.$ Also, an expansion for
the centrifugal potential term has been performed around the point $r\approx
R_{0}$ $(x=0)$ [38], and without loss of generality we put $x\equiv \left(
r-R_{0}\right) /R_{0}$ at the end of our calculations.

The empirical values found by Perey \textit{et al}. are given as $%
R_{0}=1.285 $ $fm$ and $a=0.65$ $fm$ [44]. In addition, the following WS
potential strength parameter is $V_{0}\approx 40.5+0.13A_{0}$ $MeV$ in the
non-relativistic limit$.$ Here, $A_{0}$ is the atomic mass number of target
nucleus and is defined through $R_{0}=r_{0}A_{0}^{1/3}.$ On the other hand,
the associated upper-spinor component of the wave functions is%
\begin{equation*}
F_{nl}(z)=\mathcal{N}_{n}z^{\varepsilon _{nl}}(1-qz)^{\frac{1}{2}\left(
1+\xi _{1}\right) }P_{n}^{(2\varepsilon _{nl},\xi _{1})}(1-2qz)
\end{equation*}%
\begin{equation}
=N_{n}z^{\varepsilon _{nl}}(1-qz)^{\frac{1}{2}\left( 1+\xi _{1}\right) }%
\begin{array}{c}
_{2}F_{1}%
\end{array}%
\left( -n,1+2\varepsilon _{nl}+\xi _{1}+n;2\varepsilon _{nl}+1;qz\right) ,%
\text{ }
\end{equation}%
where
\begin{equation}
\varepsilon _{nl}=a\sqrt{-\frac{2\mu }{\hbar ^{2}}E_{nl}+\frac{l(l+1)}{%
R_{0}^{2}}D_{0}}>0,\text{ }\xi _{1}=\sqrt{1+\frac{4l(l+1)a^{2}}{%
q^{2}R_{0}^{2}}D_{2}},\text{ }l=0,1,2,\cdots ,
\end{equation}%
where $E_{nl}<\frac{l(l+1)\hbar ^{2}}{2\mu R_{0}^{2}}D_{0}$ is the essential
condition for existing bound-states. As a numerical example, we impose
appropriate values for the parameters in Eq. (51) to calculate the bound
state energies of the spin symmetry generalized WS potential for special
case $\kappa =-1$ and using $\hbar =c=1.$ The results obtained by using the
following parameters $V_{0}=2.2,$ $m=15,$ $A=-5$ and $a=1.425$ are given in
Table 2. The condition of existing bound states is the $A<0.$ When the $%
A\geq 0,$ there are no bound states in the limit of exact spin symmetry. For
$q=\pm 1,$ there is only one attractive bound state $E_{0,-1}\approx -\left(
A+m\right) .$

\section{Conclusions}

We have discussed the approximate bound state solutions of the Dirac
equation for the generalized WS potential with any arbitrary spin-orbit $%
\kappa $ state under the conditions of the spin (pseudospin) symmetry $V-S=A$
($V+S=A$) by means of the NU method combined with the approximation for the
centrifugal term. By setting $V+S$ ($V-S$) to the spherically symmetric WS
potential, we have derived the solutions of the Dirac equation for the
relativistic energy eigenvalues and associated two-component spinor wave
functions for arbitrary spin-orbit $\kappa $ state that provides an
approximate solution to the spin- and pseudo-spin symmetric Dirac equations.
The most stringent interesting result is that the present spin and
pseudo-spin symmetric can be easily reduced to the KG solution once $%
S(r)=V(r)$ and $S(r)=-V(r)$ (\textit{i.e}., $A=0$)$,$ respectively. The
non-relativistic limits of our solution are obtained by imposing appropriate
transformations. The resulting solutions of the wave functions are being
expressed in terms of the Jacobi polynomials. If we choose the spin-orbit
quantum number $\kappa =-1$ ($\kappa =1$) for spin (pseudospin) symmetry$,$
the problem reduces to the exact $s(\widetilde{s})$-wave Dirac solution. The
generalized Hulth\'{e}n potential bound state solutions are simply derived
when letting $q\rightarrow -q$. We have also discussed the relation between
the non-relativistic and relativistic solutions and the possibility of
existing the bound states. It should be noted that the numerical calculation
for energy levels involved in Eq. (51) is terribly sensitive to the choice
of those parameters. In Table 2, we choose $E_{n,-1}^{1}$ as the physical
solution for the transcendental equation (51).

At the end, the solutions that constitute the main results regarding the
energy equations (31) and (49) for the spin and pseudospin symmetry,
respectively, may have some interesting applications in many areas in
physics. For example, the work is helpful to understand spectroscopy with
high field physics and useful to understand the nuclear properties like
nuclear scattering systems [45]. In addition, the present results play an
essential role in microscopic physics, since it can be used to describe the
interaction of a nucleon with a heavy nucleus [21,22]. In the
non-relativistic limits, the energy eigenvalues, Eq. (58), is physical and
is in good agreement with the results obtained previously by other methods
and works [22].

\acknowledgments We wish to thank the anonymous referees for their
invaluable suggestions and comments that have improved the paper greatly. We
are also grateful for the partial support provided by the Scientific and
Technological Research Council of Turkey (T\"{U}B\.{I}TAK).

\appendix

\section{Parametric Generalization Version of the NU Method}

We complement the theoretical formulation of the NU method in presenting the
essential polynomials, energy equation and wave functions together with
their relevant constants as follows [46].

(i) The key polynomials:
\begin{equation}
\pi (z)=c_{4}+c_{5}z-\left[ \left( \sqrt{c_{9}}+c_{3}\sqrt{c_{8}}\right) z-%
\sqrt{c_{8}}\right] ,
\end{equation}%
\begin{equation}
k=-\left( c_{7}+2c_{3}c_{8}\right) -2\sqrt{c_{8}c_{9}}.
\end{equation}%
\begin{equation}
\tau (z)=1-\left( c_{2}-2c_{5}\right) z-2\left[ \left( \sqrt{c_{9}}+c_{3}%
\sqrt{c_{8}}\right) z-\sqrt{c_{8}}\right] ,
\end{equation}%
\begin{equation}
\tau ^{\prime }(z)=-2c_{3}-2\left( \sqrt{c_{9}}+c_{3}\sqrt{c_{8}}\right) <0,
\end{equation}%
(ii) The energy equation:%
\begin{equation}
\left( c_{2}-c_{3}\right) n+c_{3}n^{2}-\left( 2n+1\right) c_{5}+\left(
2n+1\right) \left( \sqrt{c_{9}}+c_{3}\sqrt{c_{8}}\right) +c_{7}+2c_{3}c_{8}+2%
\sqrt{c_{8}c_{9}}=0.
\end{equation}%
(iii) The wave functions:%
\begin{equation}
\rho (z)=z^{c_{10}}(1-c_{3}z)^{c_{11}},
\end{equation}%
\begin{equation}
\phi (z)=z^{c_{12}}(1-c_{3}z)^{c_{13}},\text{ }c_{12}>0,\text{ }c_{13}>0,
\end{equation}%
\begin{equation}
y_{n}(z)=P_{n}^{\left( c_{10},c_{11}\right) }(1-2c_{3}z),\text{ }c_{10}>-1,%
\text{ }c_{11}>-1,
\end{equation}%
\begin{equation}
u(z)=\mathcal{N}_{n}z^{c_{12}}(1-c_{3}z)^{c_{13}}P_{n}^{\left(
c_{10},c_{11}\right) }(1-2c_{3}z),
\end{equation}%
where $P_{n}^{\left( \mu ,\nu \right) }(x),$ $\mu >-1,\nu >-1$ and $x\in
\lbrack -1,1]$ are the Jacobi polynomials with%
\begin{equation}
P_{n}^{\left( \alpha ,\beta \right) }(1-2s)=\frac{\left( \alpha +1\right)
_{n}}{n!}%
\begin{array}{c}
_{2}F_{1}%
\end{array}%
\left( -n,1+\alpha +\beta +n;\alpha +1;s\right) ,
\end{equation}%
and $\mathcal{N}_{n}$ is a normalization constants. Also, the above wave
functions can be expressed in terms of the hypergeometric function as%
\begin{equation}
u_{n\kappa }(z)=\mathcal{N}_{n\kappa }z^{c_{12}}(1-c_{3}z)^{c_{13}}%
\begin{array}{c}
_{2}F_{1}%
\end{array}%
\left( -n,1+c_{10}+c_{11}+n;c_{10}+1;c_{3}z\right) ,
\end{equation}%
where $c_{12}>0,$ $c_{13}>0$ and $z\in \left[ 0,1/c_{3}\right] .$

(iv) The relevant constants:%
\begin{equation*}
c_{4}=\frac{1}{2}\left( 1-c_{1}\right) ,\text{ }c_{5}=\frac{1}{2}\left(
c_{2}-2c_{3}\right) ,\text{ }c_{6}=c_{5}^{2}+B_{1},
\end{equation*}%
\begin{equation*}
\text{ }c_{7}=2c_{4}c_{5}-B_{2},\text{ }c_{8}=c_{4}^{2}+B_{3},\text{ }%
c_{9}=c_{3}\left( c_{7}+c_{3}c_{8}\right) +c_{6},
\end{equation*}%
\begin{equation*}
c_{10}=c_{1}+2c_{4}+2\sqrt{c_{8}}-1>-1,\text{ }c_{11}=1-c_{1}-2c_{4}+\frac{2%
}{c_{3}}\sqrt{c_{9}}>-1,
\end{equation*}%
\begin{equation}
c_{12}=c_{4}+\sqrt{c_{8}}>0,\text{ }c_{13}=-c_{4}+\frac{1}{c_{3}}\left(
\sqrt{c_{9}}-c_{5}\right) >0.
\end{equation}%
$\label{appendix}$

\section{Normalization of the radial wave function}

In order to find the normalization constants $\mathcal{N}_{n\kappa }$, we
start by writting the normalization condition:%
\begin{equation}
a\mathcal{N}_{n\kappa }^{2}\int_{0}^{1}z^{2\varepsilon _{n\kappa
}-1}(1-z)^{\xi +1}\left[ P_{n}^{(2\varepsilon _{n\kappa },\xi )}(1-2z)\right]
^{2}dz=1.
\end{equation}%
where $q=1.$ Unfortunately, there is no formula available to calculate this
key integration. Neveretheless, we can find the explicit normalization
constant $\mathcal{N}_{n\kappa }.$ For this purpose, it is not difficult to
obtain the results of the above integral by using the following formulas
[46,47]%
\begin{equation}
\int_{0}^{1}\left( 1-s\right) ^{\mu -1}s^{\nu -1}%
\begin{array}{c}
_{2}F_{1}%
\end{array}%
\left( \alpha ,\beta ;\gamma ;as\right) dz=\frac{\Gamma (\mu )\Gamma (\nu )}{%
\Gamma (\mu +\nu )}%
\begin{array}{c}
_{3}F_{2}%
\end{array}%
\left( \nu ,\alpha ,\beta ;\mu +\nu ;\gamma ;a\right) ,\text{ }\mu >-1,\text{
}\nu >-1,
\end{equation}%
and $%
\begin{array}{c}
_{2}F_{1}%
\end{array}%
\left( a,b;c;z\right) =\frac{\Gamma (c)}{\Gamma (a)\Gamma (b)}%
\sum_{p=0}^{\infty }\frac{\Gamma (a+p)\Gamma (b+p)}{\Gamma (c+p)}%
\frac{z^{p}}{p!}.$ Following Ref. [46], we calculate the
normalization
constants:%
\begin{equation}
\mathcal{N}_{n\kappa }=\left[ \frac{a\Gamma (2\varepsilon _{n\kappa
}+1)\Gamma (\xi +2)}{\Gamma (n)}\sum_{m=0}^{\infty }\frac{%
(-1)^{m}\left( 1+n+2\varepsilon _{n\kappa }+\xi )\right) _{m}\Gamma (n+m)}{%
m!\left( m+2\varepsilon _{n\kappa }\right) !\Gamma \left( m+2\varepsilon
_{n\kappa }+\xi +2\right) }C_{n\kappa }\right] ^{-1/2}\text{ }
\end{equation}%
where
\begin{equation}
C_{n\kappa }=%
\begin{array}{c}
_{3}F_{2}%
\end{array}%
\left( 2\varepsilon _{n\kappa }+m,-n,n+1+\xi +2\varepsilon _{n\kappa
};m+2\varepsilon _{n\kappa }+\xi +2;1+2\varepsilon _{n\kappa };1\right) ,
\end{equation}%
Furthermore, the normalization constants for the $s$-wave can be also found
as%
\begin{equation}
\\mathcal{N}_{n\kappa }=\left[ \frac{a\Gamma (2\varepsilon _{n\kappa
}+1)\Gamma (\xi +2)}{\Gamma (n)}\sum_{m=0}^{\infty }\frac{%
(-1)^{m}\left( 1+n+2\varepsilon _{n\kappa }+\xi )\right) _{m}\Gamma (n+m)}{%
m!\left( m+2\varepsilon _{n\kappa }\right) !\Gamma \left(
m+2\varepsilon _{n\kappa }+\xi +2\right) }C_{n\kappa }\right]
^{-1/2}\text{ }
\end{equation}%
where
\begin{equation}
C_{n}=%
\begin{array}{c}
_{3}F_{2}%
\end{array}%
\left( 2\varepsilon _{n}+m,-n,n+2\varepsilon _{n}+2;m+2\varepsilon
_{n}+3;1+2\varepsilon _{n};1\right) ,
\end{equation}%
and $\varepsilon _{n}$ is given in Eq. (53).

\newpage

{\normalsize 
}

\newpage

\bigskip

\baselineskip= 2\baselineskip
\bigskip \newpage {\normalsize 
}

\baselineskip= 2\baselineskip
\bigskip \newpage
\begin{table}[tbp]
\caption{The specific values for the parametric constants necessary for the
present potential.}%
\begin{tabular}{llll}
\tableline Constant & Value & Constant & Value \\
\tableline$c_{1}$ & 1 & $c_{2}$ & $q$ \\
$c_{3}$ & $q$ & c$_{4}$ & $0$ \\
$c_{5}$ & $-\frac{q}{2}$ & $c_{6}$ & $\frac{1}{4}\left( q^{2}+4\beta
_{1}\right) $ \\
$c_{7}$ & $-\beta _{2}$ & $c_{8}$ & $\varepsilon _{n\kappa }^{2}$ \\
$c_{9}$ & $\left( \frac{q}{2}\right) ^{2}\xi ^{2}$ & $c_{10}$ & $%
2\varepsilon _{n\kappa }$ \\
$c_{11}$ & $\xi $ & $c_{12}$ & $\varepsilon _{n\kappa }$ \\
$c_{13}$ & $\frac{1}{2}\left( 1+\xi \right) $ & $B_{1}$ & $\beta _{1}$ \\
$B_{2}$ & $\beta _{2}$ & $B_{3}$ & $\varepsilon _{n\kappa }^{2}$ \\
\tableline &  &  &  \\
&  &  &
\end{tabular}%
\end{table}

\bigskip
\begin{table}[tbp]
\caption{The bound state energy levels $E_{n\protect\kappa }$ for the
special case $\protect\kappa =-1.$}%
\begin{tabular}{lllllllll}
\tableline & $q=1$ &  & $q=2$ &  & $q=-1$ &  & $q=-2$ &  \\
$n$ & $E_{n,-1}^{1}$ & $E_{n,-1}^{2}$ & $E_{n,-1}^{1}$ & $E_{n,-1}^{2}$ & $%
E_{n,-1}^{1}$ & $E_{n,-1}^{2}$ & $E_{n,-1}^{1}$ & $E_{n,-1}^{2}$ \\
\tableline$0$ & $-10.197602$ & $-19.996367$ & $1.337420$ & $-19.996426$ & $%
-9.561001$ & $-19.996589$ & $2.018956$ & $-19.996536$ \\
$1$ & $0.985745$ & $-19.985463$ & $9.849615$ & $-19.985698$ & $2.349482$ & $%
-19.986350$ & $10.803623$ & $-19.986141$ \\
$2$ & $6.597152$ & $-19.967274$ & $12.216763$ & $-19.967804$ & $8.327343$ & $%
-19.969273$ & $13.247481$ & $-19.968802$ \\
$3$ & $9.328912$ & $-19.941777$ & $13.119033$ & $-19.942722$ & $11.239603$ &
$-19.945338$ & $14.180142$ & $-19.944500$ \\
$4$ & $10.775321$ & $-19.908939$ & $13.534261$ & $-19.910418$ & $12.784029$
& $-19.914518$ & $14.610668$ & $-19.913204$ \\
$5$ & $11.602814$ & $-19.868715$ & $13.742899$ & $-19.870854$ & $13.670315$
& $-19.876777$ & $14.828469$ & $-19.874879$ \\
$6$ & $12.102341$ & $-19.821054$ & $13.847458$ & $-19.823977$ & $14.208309$
& $-19.832071$ & $14.939340$ & $-19.829478$ \\
$7$ & $12.412538$ & $-19.765892$ & $13.892224$ & $-19.769728$ & $14.545655$
& $-19.780348$ & $14.988985$ & $-19.776946$ \\
$8$ & $12.605105$ & $-19.703154$ & $13.898750$ & $-19.708035$ & $14.758727$
& $-19.721546$ & $14.999654$ & $-19.717219$ \\
$9$ & $12.719786$ & $-19.632753$ & $13.878495$ & $-19.638818$ & $14.889869$
& $-19.655595$ & $14.983185$ & $-19.650223$ \\
$10$ & $12.779911$ & $-19.554593$ & $13.837978$ & $-19.561982$ & $14.963949$
& $-19.582415$ & $14.946314$ & $-19.575874$ \\
\tableline &  &  &  &  &  &  &  &
\end{tabular}%
\end{table}

\bigskip

\bigskip


\begin{thebibliography}{99}
\bibitem{1} L.I. Schiff, Quantum Mechanics, third ed., McGraw-Hill, New
York, 1995.

\bibitem{2} L.D. Landau and E.M. Lifshitz, Quantum Mechanics,
Non-Relativistic Theory, third ed., Pergamon, New York, 1977.

\bibitem{3} S.H. Dong, Factorization Method in Quantum Mechanics, Springer,
2007.

\bibitem{4} X.-C. Zhang, Q.-W. Liu, C.-S. Jia and L.-Z. Wang, Phys. Lett. A
\textbf{340}, 59 (2005); I.C. Wang and C.Y. Wong, Phys. Rev. D \textbf{34},
348 (1988).

\bibitem{5} A. Sinha and P. Roy, Mod. Phys. Lett. A \textbf{20}, 2377
(2005); C.S. Jia and A. de S. Dutra, J. Phys. A: Math. Gen. \textbf{39},
11877 (2006); A. de S. Dutra and C.S. Jia, Phys. Lett. A \textbf{352}, 484
(2006).

\bibitem{6} S.M. Ikhdair, Int. J. Mod. Phys. C \textbf{20 }(1), 25 (2009).

\bibitem{7} H. E\u{g}rifes and R. Sever, Phys. Lett. A \textbf{344}, 117
(2005).

\bibitem{8} F. Dominguez-Adame, Phys. Lett. A \textbf{136}, 175 (1989); F.
Dominguez-Adame and A. Rodriguez, Phys. Lett. A \textbf{198}, 275 (1995).

\bibitem{9} N.A. Rao and B.A. Kagali, Phys. Lett. A \textbf{296}, 192 (2002).

\bibitem{10} L-Z Yi, Y-F Diao, J-Y Liu and C-S Jia, Phys. Lett. A \textbf{333%
}, 212 (2004).

\bibitem{11} S.M. Ikhdair and R. Sever, Int. J. Mod. Phys. E \textbf{17},
1107 (2008).

\bibitem{12} S.M. Ikhdair and R. Sever, Ann. Phys. (Berlin) \textbf{16}, 218
(2007).

\bibitem{13} S.M. Ikhdair and R. Sever, Int. J. Theor. Phys. \textbf{46},
1643; 2384 (2007); S.M. Ikhdair and R. Sever, J. Math. Chem. \textbf{42}
(3), 461 (2007).

\bibitem{14} S.M. Ikhdair, Chin. J. Phys. \textbf{46}, 291 (2008); S.M.
Ikhdair and R. Sever, Int. J. Mod. Phys. C \textbf{18}, 1571 (2007);\textit{%
\ }Int. J. Mod. Phys.\textit{\ }C \textbf{19}, 221 (2008); Cent. Eur. J.
Phys.\textbf{\ 6}, 685, 697 (2008).

\bibitem{15} Y.F. Cheng and T.Q. Dai, Phys. Scr. \textbf{75}, 274 (2007).

\bibitem{16} C. Berkdemir, A. Berkdemir and J. Han, Chem. Phys. Lett.
\textbf{417}, 326 (2006).

\bibitem{17} S.M. Ikhdair and R. Sever, J. Mol. Struc.:Theochem \textbf{806}%
, 155 (2007); J. Mol. Struc.:Theochem \textbf{809}, 103 (2007); J. Mol.
Struc.:Theochem \textbf{855}, 13 (2008); J. Math. Chem. \textbf{41}, 329
(2007); J. Math. Chem. \textbf{41}, 343 (2007).

\bibitem{18} S.M. Ikhdair and R. Sever, Cent. Eur. J. Phys. \textbf{6}, 141
(2008); Cent. E. J. Phys. \textbf{5}, 516 (2007).

\bibitem{19} S.M. Ikhdair and R. Sever, Ann. Phys. (Berlin) \textbf{17}, 897
(2008); Ann. Phys. (Berlin) \textbf{18 }(4), 189 (2009).

\bibitem{20} S.M. Ikhdair and R. Sever, Phys. Scr. \textbf{79 }(3), 035002
(2009); S.M. Ikhdair, Eur. Phys. J. A \textbf{39 }(3), 307 (2009); S.M.
Ikhdair and R. Sever, Int. J. Mod. Phys. C \textbf{19}, 1425 (2008); S.M.
Ikhdair and R. Sever, Int. J. Mod. Phys. C \textbf{20 }(3), 361 (2009); S.M.
Ikhdair and R. Sever, J. Math. Chem. \textbf{45 }(4), 1137 (2009).

\bibitem{21} M.E. Grypeos and B.A. Kotsos, J. Phys. B:At. Mol. Opt. Phys.
\textbf{29}, L473 (1996); B.A. Kotsos and M. Grypeos, Physica B \textbf{229}%
, 173 (1997); S. Fl\"{u}gge, Practical Quantum Mechanics, Springer-Verlag,
Berlin, 1974.

\bibitem{22} C. Berkdemir, A. Berkdemir and R. Sever, Phys. Rev. C \textbf{72%
}, 027001 (2005); Editorial Note, Phys. Rev. C 74, 039902 (E) (2006); ibid.,
J. Phys. A: Math. Gen. \textbf{39},\textbf{\ }13455 (2006).

\bibitem{23} P. Kennedy, J. Phys. A \textbf{35}, 689 (2002).

\bibitem{24} J.-Y. Guo and Z.-Q. Sheng, Phys. Lett. A \textbf{338}, 90
(2005).

\bibitem{25} A.D. Alhaidari, Phys. Rev. Lett. \textbf{87}, 210405 (2001);
A.D. Alhaidari, Phys. Rev. Lett. \textbf{88}, 189901 (2002).

\bibitem{26} A.D. Alhaidari, J. Phys. A \textbf{34}, 9827 (2001); A.D.
Alhaidari, J. Phys. A \textbf{35}, 6207 (2002).

\bibitem{27} A.D. Alhaidari, Phys. Lett. A \textbf{322}, 72 (2004); A.D.
Alhaidari, Phys. Lett. A \textbf{326}, 58 (2004).

\bibitem{28} J.-Y. Guo, X.-Z. Fang and F.X. Xu, Phys. Rev. A \textbf{66},
062105 (2002).

\bibitem{29} J.-Y. Guo, J. Meng and F.X. Xu, Chin. Phys. Lett. \textbf{20},
602 (2003).

\bibitem{30} J.N. Ginocchio, Phys. Rev. \textbf{69}, 034318 (2004); J.N.
Ginocchio, Phys. Rev. Lett. \textbf{95}, 252501 (2005), J.N. Ginocchio,
Phys. Rep. 414, 165 (2005).

\bibitem{31} R. Lisboa and M. Malheiro, Phys. Rev. C \textbf{69}, 024319
(2004).

\bibitem{32} S.M. Ikhdair, submitted to Annals Phys. (New York) AOP-67392
(2009); ibid. AOP-67385 (2009).

\bibitem{33} S.M. Ikhdair, Eur. Phys. J. A \textbf{40 }(2)\textbf{, }143
(2009).

\bibitem{34} A.F. Nikiforov and V.B. Uvarov, Special Functions of
Mathematical Physics (Birkhauser, Basel, 1988).

\bibitem{35} G.-F. Wei and S.-H. Dong, Phys. Lett. A 373, 49 (2008); C.-S.
Jia, T. Chen and L.-G. Cui, Phys. Lett. A 373, 1621 (2009).

\bibitem{36} A.D. Alhaidari, H. Bahlouli and A. Al-Hasan, Phys. Lett. A
\textbf{349}, 87 (2006).

\bibitem{37} L.S. Costa, F.V. Prudenter, P.H. Acioli, J.J. Soares Neto and
J.D.M. Vianna, J. Phys. B \textbf{32}, 2461 (1999).

\bibitem{38} J. Lu, Phys. Scr. \textbf{72}, 349 (2005); J. Lu, H.-X. Qian,
L.-M. Li and F.-L. Liu, Chin. Phys. \textbf{14}, 2402 (2005).

\bibitem{39} C. Berkdemir and Y.-F. Cheng, Phys. Scr. \textbf{79}, 035003
(2009); S.M. Ikhdair, submitted to Phys. Scr. (2009).

\bibitem{40} A. De Souza Dutra and M. Hott, Phys. Lett. A \textbf{356}, 215
(2006).

\bibitem{41} W. Greiner, Relativistic Quantum Mechanics (Springer, Verlag,
1981).

\bibitem{42} B. G\"{o}n\"{u}l and K. K\"{o}ksal, Phys. Scr. \textbf{76}, 565
(2007).

\bibitem{43} R.L. Greene and C. Aldrich, Phys. Rev. A \textbf{14}, 2363
(1976); B. G\"{o}n\"{u}l and \.{I}. Zorba, Phys. Lett. A \textbf{269}, 83
(2000).

\bibitem{44} C.M. Perey \textit{et al}, Phys. Rev. \textbf{175}, 1460 (1968).

\bibitem{45} I. Boztosun, Phys. Rev. C \textbf{66}, 024610 (2002).

\bibitem{46} S.M. Ikhdair, Int. J. Mod. Phys. C \textbf{20} (10) (2009)
[arXiv:0905.2867]; S.M. Ikhdair, Chem. Phys. \textbf{361} (1-3) (2009)
[DOI:10.1016/j.chemphys.2009.04.023] [arXiv:0904.4366].

\bibitem{47} M. Abamowitz and I.A. Stegun, Handbook of Mathematical
Functions (Dover, New York, 1970); S. Gradshteyn and I.M. Ryzhik, Tables of
Integrals, Series and Products, 5th edn. (Academic, New York, 1994).
\end{thebibliography}
\end{document}